\input harvmac 
\input pictex
\overfullrule=0pt
\def\half{{\textstyle{1\over 2}}}

\def\neqno#1{\eqnn{#1} \eqno #1}
\def\alt{\mathop{\raise.3ex\hbox{$<$\kern-.75em\lower1ex\hbox{$\sim$}}} }
\font\bbb=msbm10
\font\bbbs=msbm10 at 9pt

\Title{\vbox{ \hfill hep-th/0110322 \vskip-3pt
\hfill UTPT-01-10\vskip-3pt \hfill CMU-HEP-01-10}} {\vbox{
\centerline{Matter in a Warped and Oscillating Background}  }}
\centerline{Hael Collins\footnote{$^\dagger$}{{\tt 
hael@physics.utoronto.ca}} }
\medskip
\vbox{\it \centerline{Department of Physics} \vskip-2pt
\centerline{Carnegie Mellon University} \vskip-2pt
\centerline{Pittsburgh, PA\ \ 15213}}
\bigskip

\centerline{Bob Holdom\footnote{$^\ddagger$}{{\tt bob.holdom@utoronto.ca}} }
\medskip
\vbox{\it \centerline{Department of Physics} \vskip-2pt
\centerline{University of Toronto} \vskip-2pt
\centerline{Toronto, Ontario  M5S 1A7, Canada}}

\bigskip
\bigskip
\centerline{Abstract}

\medskip 
{\baselineskip=10pt\centerline{\vbox{\hsize=.8\hsize\ninepoint\noindent  
We examine the role of matter in an oscillating background with a warped,
compact extra dimension.  This background is compatible with an
$S^1/\hbox{\bbbs Z}_2$ orbifold structure which allows chiral fermions to be
included in the scenario.  When the background oscillates rapidly, the leading
coupling of these oscillations is to gauge fields rather than fermions.  If
the decay of these oscillations were to occur today, it could provide an
alternative mechanism for generating the ultra high energy cosmic rays.  }}}

\Date{October, 2001}
\baselineskip=12pt

\newsec{Introduction.}

A remarkable property of actions that generalize the Einstein-Hilbert action
for gravity is that they admit metrics with a periodic dependence on one or
more of the coordinates.  The simplest such example occurs for an action
composed of a general set of curvature invariants, with up to four derivatives
of the metric, and scalar field.  In $4+1$ dimensions a metric periodic in one
of the spatial dimensions provides a naturally compact space---by identifying
the compactification length with the period---without any singularities or
discontinuities in the metric 
\ref\warped{H.~Collins and B.~Holdom,
Phys.\ Rev.\ D {\bf 63}, 084020 (2001) [hep-th/0009127].}.   
Moreover, these periodic solutions exist without the need for finely tuning
any of the parameters in the action.  The coefficients of the four-derivative
terms determine an essentially unique compactification size.

Similarly, we can search for metrics that oscillate both in the fifth
dimension and in time 
\ref\osc{H.~Collins and B.~Holdom,
hep-th/0107042.}.   
Although these metrics do not describe the most general isotropic vacuum
solutions for the action, they could have intriguing consequences for the
early and the late evolution of the universe.  In the very early universe, we
shall see that the rapid decay of the oscillating background into gauge fields
provides an example of how a universe with a Planck-scale time dependence can
relax into one that can admit a realistic, slow evolution.  Moreover, if some
slowly relaxing, but rapid, oscillation persisted or arose recently, then it
could contribute a flux of ultra high energy cosmic rays.  To be successful,
the theory would still require a mechanism to maintain these oscillations at a
satisfactory rate of decay today.

In this article we shall explore the effect of including fermion and gauge
fields in these oscillating backgrounds.  Even for a purely static background,
introducing chiral fermions requires that the compact extra dimension should
have the topology of an $S^1/${\bbb Z}$_2$ orbifold.  The metric and the
scalar field that supports the compact geometry must be respectively even and
odd under this {\bbb Z}$_2$.  When the metric additionally oscillates rapidly
in time, the coupling of these fields with gravity leads to a potential decay
channel for the oscillations.  We shall show by expanding a fermion and an
Abelian gauge field in Kaluza-Klein modes that to leading order the fermion
zero mass mode does not couple to the oscillating component of the metric. 
The preferred channel for the relaxation of the metric is therefore into gauge
fields.

In the following section we review the features of oscillating metrics in
$4+1$ dimensions.  Section 3 describes how to place the theory on an orbifold
to allow for chiral fermions in a warped, static background.  Section 4
examines the coupling of the time-dependent components of the metric with the
Kaluza-Klein modes of a fermion and an Abelian gauge field placed in an
oscillating, compact background.  Finally, in section 5, we study the
phenomenological signature in ultra high energy cosmic rays of a very small,
but rapid, oscillation today.

\newsec{Oscillating metrics.}

At distances approaching the Planck length, the usual Einstein-Hilbert action
should be supplemented by higher derivative curvature invariants.  For
example, an action with up to four derivatives of the metric and a scalar
field has the form\foot{Our convention for the signature of the metric is
$(-,+,+,+,+)$ while the Riemann curvature tensor is defined by $-R^a_{\ bcd}
\equiv \partial_d \Gamma^a_{bc} - \partial_c \Gamma^a_{bd} +
\Gamma^a_{ed}\Gamma^e_{bc} - \Gamma^a_{ec}\Gamma^e_{bd}$.  Here the indices
$a,b,\ldots = 0,1,2,3,y$ range over all coordinates while $\mu,\nu\, \ldots =
0,1,2,3$ label the ordinary space-time coordinates.  }
$$\eqalign{S 
&= M_5^3 \int d^4xdy\, \sqrt{-g}\, \left[ -2\Lambda + R +
a\, R^2 + b\, R_{ab}R^{ab} + c\, R_{abcd}R^{abcd} \right] \cr 
&\quad +
M_5^3 \int d^4xdy\, \sqrt{-g}\, \left[ - {\textstyle{1\over 2}}
\nabla_a\phi\nabla^a\phi + \Delta{\cal L}_\phi \right] + \cdots . \cr}
\neqno\action$$
$M_5$ and $\Lambda$ denote the Planck mass and cosmological constant
respectively.  This metric admits smooth, non-singular metrics that are
compact in the fifth dimension, 
$$ds^2 = g_{ab}\, dx^a dx^b = e^{A(y)}\, \eta_{\mu\nu}\,  dx^\mu dx^\nu 
+ dy^2  \neqno\Ametric$$
where the exponent $A(y)$ is a periodic function, when $\Delta{\cal L}_\phi$
is either an interaction, such as $(\nabla_a\phi\nabla^a\phi)^2$ $\warped$, or
a Casimir term 
\ref\warpedRS{H.~Collins and B.~Holdom,
Phys.\ Rev.\ D {\bf 64}, 064003 (2001) [hep-ph/0103103].}  
with differing values its components in the large ordinary $x^\mu$ and the
compact $y$ directions.  In either case, the existence of such solutions does
not require finely tuning any of the parameters in the action.  The effects of
the $R^2$ terms can be conveniently parameterized by 
$$\mu = 16a+5b+4c,\quad
\lambda = 5a+b+{\textstyle{1\over 2}}c \quad\hbox{and}\quad
\nu = 3a+b+c . \neqno\mulambdanu$$
$\lambda$ and $\nu$, in particular, represent the coefficients of a
Gauss-Bonnet and a squared Weyl tensor respectively.

While the static metrics are adequate for studying a theory with a flat $3+1$
dimensional long distance limit, models that are to include a realistic
cosmology should evolve in time as well,
$$ds^2 = - e^{A(y)}\, dt^2 + e^{A(y)}e^{B(t)}\, \delta_{ij}\, dx^i dx^j 
+ e^{C(t)}\, dy^2 ; \neqno\oldmetric$$
here we have still assumed an isotropy in the three large spatial dimensions
but have allowed a different evolution in the compact dimension.  

We shall principally consider a universe that oscillates rapidly so that the
time dependence and the $y$-dependence, which fixes the size of the compact
dimension, are on a similar footing.  The fourth-order, two-variable
differential equations that result from varying the action $\action$ are more
difficult to solve than the static case $\Ametric$, which can be solved
numerically, but fortunately it is possible to build a solution order by order
in a small amplitude expansion.  To first order in $\epsilon_y,\epsilon_t\ll
1$, we find $\osc$
$$\eqalign{
A(y) &= \epsilon_y \cos(\omega_y y) + \cdots \cr
B(t) &= \epsilon_t\cos(\omega_t t) + \cdots 
\qquad
C(t) = - 3 \epsilon_t\cos(\omega_t t) + \cdots \cr}
\neqno\ABCone$$
with
$$\omega_y = \sqrt{{-3\over\mu}} \quad\hbox{and}\quad
\omega_t = {1\over\sqrt{3\mu-16\nu}} \neqno\ABComegas$$
so that a periodic solution exists when $\mu<0$ and $\nu < {3\over 16}
\mu$---no fine-tuning is needed.  In $\ABComegas$ we see explicitly that the
higher derivative terms in the action determine the size of the extra
dimension and the oscillation frequency.  Proceeding to the next order in the
small $\epsilon_y, \epsilon_t$ expansion $\osc$ introduces some corrections to
each of $A(y)$, $B(t)$ and $C(t)$, which depend on {\it both\/} $t$ and $y$,
and more importantly relates the small amplitudes to the size of the
cosmological constant:  
$$\Lambda = {\textstyle{3\over 4}} \left( \omega_y^2 \epsilon_y^2 
+ \omega_t^2 \epsilon_t^2 \right) . \neqno\cosconstant $$
The relative size of the $\epsilon_y$ and $\epsilon_t$ in this solution is
undetermined.  This relation allows us to state more precisely the regime in
which the small amplitude expansion exists.  When $\mu$ and $\nu$ have a
natural size, ${\cal O}(M_5^{-2})$, then a small amplitude translates to a
small cosmological constant, $M_5\Lambda^{-1/2} \gg 1$.  Note however that the
existence of periodic solutions---beyond the small amplitude regime---does not
require a small cosmological constant as was shown numerically in $\warped$.

In the large $3+1$ dimensions, once we have substituted the leading effects
from the $y$-dependent scalar field, we discover that the density $\rho$ and
the pressure $p$ for the non-compact spatial dimensions are 
$$\eqalign{
\rho &= {\textstyle{3\over 4}} \omega_t^2\epsilon_t^2
+ {\textstyle{1\over 4}}\dot\phi^2 + \cdots 
\approx {\textstyle{3\over 2}} \omega_t^2\epsilon_t^2 > 0 \cr
p &= - {\textstyle{3\over 4}} \omega_t^2\epsilon_t^2 
+ {\textstyle{1\over 4}}\dot\phi^2 + \cdots \approx 0 \cr} 
\neqno\rhoandp$$
so that ${3\over 4} \omega_t^2\epsilon_t^2$ resembles an effective
cosmological constant.  The fundamental cosmological constant that appears in
the action effects the warping of the extra dimension, as the pressure in that
dimension indicates, $p_y = -2 \Lambda + {3\over 2} \omega_t^2\epsilon_t^2$.

We find that the small amplitude expansion described above cannot be extended
to the case of a de Sitter expansion $B(t) = kt + \cdots$ that coexists with
the oscillations, if the rate of expansion is to be comparable to the leading
oscillating terms in the metric, {\it i.e.\/} $k\sim {\cal O}(M_5\epsilon_t)$. 
In general, the dynamics that links the short scale oscillations with the
large scale evolution of the universe is complicated and deserves further
study.

If the amplitude of the oscillations starts with a value large compared to any
expansion rate, the amplitude must eventually decrease sufficiently so as to
allow a realistic cosmology.  Given that the vacuum energy density in
$\rhoandp$ is positive, a natural mechanism for its relaxation is through the
decay of the rapid oscillations into energetic particles, analogous to the
decay of the inflaton in inflation.  In section 4, we shall study how fermions
and gauge fields couple to the oscillating functions, $B(t)$ and $C(t)$, thus
providing an explicit decay mechanism.  

Note that the presence of the extra dimension is necessary for this decay.  In
$3+1$ dimensions the metric would not contain a $C(t)$ term and would
therefore be conformally flat; then the oscillation could not decay into
conformally coupled fields.

\newsec{Chiral fermions.}

In order to recover the standard model fields, the theory must contain chiral
fermions.  A difficulty arises for the usual Kaluza-Klein compactification
with only one extra dimension since the Lorentz symmetry group $SO(4,1)$ has
only one, {\it non-chiral\/}, spin-${1\over 2}$ representation.  Chiral spinor
representations do arise when we break the full $SO(4,1)$ symmetry group, for
example by placing the theory on an $S^1/\hbox{\bbb Z}_2$ orbifold in the
fifth dimension %
\ref\chiral{
K.~R.~Dienes, E.~Dudas and T.~Gherghetta,
Nucl.\ Phys.\ B {\bf 537}, 47 (1999) [hep-ph/9806292] 
and 
H.~Cheng, B.~A.~Dobrescu and C.~T.~Hill,
Nucl.\ Phys.\ B {\bf 589}, 249 (2000) [hep-ph/9912343].}.  

We can similarly introduce an orbifold into the warped Kaluza-Klein picture
with a static metric, $\Ametric$.  The field equations for the action
$\action$ that determine $A(y)$ and $\phi(y)$ do not depend explicitly on $y$
so we are free to translate an extremum of $A(y)$ to $y=0$.  If the period of
$A(y)$ is $y_c$ then we can compactify the space by restricting $y\in
[-{1\over 2}y_c,{1\over 2}y_c]$ and identifying the endpoints.  The solutions
for $A(y)$ found numerically in $\warped$ and analytically in a small
amplitude expansion in $\osc$ are manifestly even under $y\to -y$ so the
background metric respects an $SO(3,1)\times\hbox{\bbb Z}_2$ symmetry.

This {\bbb Z}$_2$ invariance allows us to introduce an orbifold geometry in
the extra dimension by identifying $y$ with $-y$.  In order to define a theory
consistently on this orbifold, the fields must be odd or even under this
discrete {\bbb Z}$_2$.  As we have seen, the background metric is even and
this orbifold geometry will further constrain the allowed gravitational
excitations of this background.

The scalar field, in contrast, must be odd, $\phi(-y)=-\phi(y)$.  The reason
is that the field equations for $\action$ relate $\phi'(y)$ to an expression
that depends on $A(y)$ only through terms with even numbers of derivatives. 
Therefore, $\phi'(y)$ oscillates with the same period as $A(y)$ and is also
even under $\phi'(y)=\phi'(-y)$.  For the solutions found in $\warped$ $\osc$
$\warpedRS$, $\phi'(y)$ is everywhere positive so that after integrating we
obtain a $\phi(y)$ that increases monotonically.   Choosing the constant of
integration so that $\phi(0)=0$ produces a $\phi(y)$ that is odd.  To
accommodate the boundary values, we must further impose that $\phi(y)$ itself
assumes values only in a compact space by identifying $\phi(y_c/2)$ and
$\phi(-y_c/2)$.  In figure 1, we show explicitly a example with this geometry. 
The form for $\phi(y)$ was found numerically for $\Delta{\cal
L}_\phi=k(\nabla_a\phi\nabla^a\phi)^2$ but with the parameters chosen
arbitrarily within the region of the $\{ \Lambda, \mu, \lambda, k\}$-space
with periodic solutions in $y$.
$$\beginpicture
\setcoordinatesystem units <2.0truein,1.0truein>
\setplotarea x from -0.7 to 0.7, y from -0.8 to 0.8
\setlinear
\putrule from -0.6578 0.0 to 0.6578 0.0
\putrule from 0.0 -0.744 to 0.0 0.744
\putrule from 0.6578 -0.04 to 0.6578 0.04
\putrule from 0.50 -0.02 to 0.50 0.02
\putrule from 0.25 -0.02 to 0.25 0.02
\putrule from -0.25 -0.02 to -0.25 0.02
\putrule from -0.50 -0.02 to -0.50 0.02
\putrule from -0.6578 -0.04 to -0.6578 0.04
\putrule from -0.02 -0.744 to 0.02 -0.744
\putrule from -0.01 -0.50 to 0.01 -0.50
\putrule from -0.01 -0.25 to 0.01 -0.25
\putrule from -0.01 0.25 to 0.01 0.25
\putrule from -0.01 0.50 to 0.01 0.50
\putrule from -0.02 0.744 to 0.02 0.744
\put {$\phi(y)$} [l] at 0.7 0.744
\put {$y$} [c] at 0.7 0.15 
\put {$-\half y_c$} [c] at -0.6578 -0.12 
\put {$\half y_c$} [c] at 0.6578 -0.12 
\put {$\phi(\half y_c)$} [r] at -0.06 0.744
\put {$-\phi(\half y_c)$} [r] at -0.06 -0.744
\put {${\scriptstyle 0.25}$} [c] at 0.25 -0.1
\put {${\scriptstyle 0.5}$} [c] at 0.5 -0.1
\put {${\scriptstyle 0.25}$} [r] at -0.04 0.25
\put {${\scriptstyle 0.5}$} [r] at -0.04 0.5
%
%
\plot 
0.00 0.000  0.01 0.015  0.02 0.030  0.03 0.045  0.04 0.061  0.05 0.076
0.06 0.090  0.07 0.105  0.08 0.120  0.09 0.134  0.10 0.149  0.11 0.163
0.12 0.177  0.13 0.191  0.14 0.205  0.15 0.218  0.16 0.232  0.17 0.245
0.18 0.258  0.19 0.271  0.20 0.283  0.21 0.296  0.22 0.308  0.23 0.320
0.24 0.332  0.25 0.344  0.26 0.355  0.27 0.367  0.28 0.378  0.29 0.389
0.30 0.400  0.31 0.411  0.32 0.422  0.33 0.433  0.34 0.443  0.35 0.454
0.36 0.464  0.37 0.474  0.38 0.484  0.39 0.494  0.40 0.504  0.41 0.514
0.42 0.524  0.43 0.534  0.44 0.543  0.45 0.553  0.46 0.563  0.47 0.572
0.48 0.582  0.49 0.591  0.50 0.600  0.51 0.610  0.52 0.619  0.53 0.628
0.54 0.637  0.55 0.646  0.56 0.656  0.57 0.665  0.58 0.674  0.59 0.683
0.60 0.692  0.61 0.701  0.62 0.710  0.63 0.719  0.64 0.728  0.65 0.737
0.6578 0.744  /
\plot 
-0.00 -0.000  -0.01 -0.015  -0.02 -0.030  -0.03 -0.045  
-0.04 -0.061  -0.05 -0.076  -0.06 -0.090  -0.07 -0.105
-0.08 -0.120  -0.09 -0.134  -0.10 -0.149  -0.11 -0.163
-0.12 -0.177  -0.13 -0.191  -0.14 -0.205  -0.15 -0.218
-0.16 -0.232  -0.17 -0.245  -0.18 -0.258  -0.19 -0.271
-0.20 -0.283  -0.21 -0.296  -0.22 -0.308  -0.23 -0.320
-0.24 -0.332  -0.25 -0.344  -0.26 -0.355  -0.27 -0.367
-0.28 -0.378  -0.29 -0.389  -0.30 -0.400  -0.31 -0.411
-0.32 -0.422  -0.33 -0.433  -0.34 -0.443  -0.35 -0.454
-0.36 -0.464  -0.37 -0.474  -0.38 -0.484  -0.39 -0.494
-0.40 -0.504  -0.41 -0.514  -0.42 -0.524  -0.43 -0.534
-0.44 -0.543  -0.45 -0.553  -0.46 -0.563  -0.47 -0.572
-0.48 -0.582  -0.49 -0.591  -0.50 -0.600  -0.51 -0.610
-0.52 -0.619  -0.53 -0.628  -0.54 -0.637  -0.55 -0.646
-0.56 -0.656  -0.57 -0.665  -0.58 -0.674  -0.59 -0.683
-0.60 -0.692  -0.61 -0.701  -0.62 -0.710  -0.63 -0.719
-0.64 -0.728  -0.65 -0.737  -0.6578 -0.744    /
\endpicture$$
{\ninepoint  \baselineskip=10pt
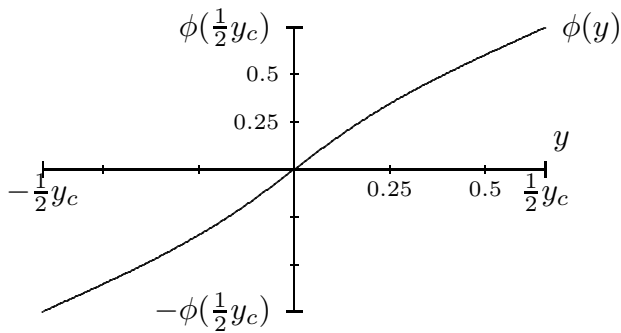
{\bf Figure 1.\/} The behavior of $\phi(y)$ on the orbifold for $\Lambda = 1$,
$\mu = 0.1$, $\lambda = 0$, and $\Delta{\cal L}_\phi = - {1\over
4}(\nabla_a\phi\nabla^a\phi)^2$.  The initial condition is
$A^{\prime\prime}(0) = 23.77364592$.  Here $y\in [-0.6578,0.6578]$ and
$\phi\in [-0.744,0.744]$.  The endpoints of each dimension are identified so
that both are compact.}
\medskip
%

In order to be compatible with this $\hbox{\bbb Z}_2$ orbifold structure, a
separate scalar field should be included if we wish also to allow the rapid
time oscillations.  This requirement follows from the leading order behavior,
$\dot\phi^2 = 3\omega_t^2\epsilon_t^2$ $\osc$; upon integration we obtain a
contribution that is even under the $\hbox{\bbb Z}_2$ symmetry which is
incompatible with the symmetry of the scalar field that supports the compact
extra dimension.

We now obtain chiral fermions through the standard construction $\chiral$ 
\ref\georgi{H.~Georgi, A.~K.~Grant and G.~Hailu,
Phys.\ Rev.\ D {\bf 63}, 064027 (2001) [hep-ph/0007350].}.   
If we choose the following boundary conditions on a five dimensional fermion,
$$\psi(x^\lambda,-y) = \gamma^5\psi(x^\lambda, y) , \neqno\fermionBC$$
then expanding in a tower of Kaluza-Klein modes, we have a chiral zero mass
mode,
$$\psi^{(0)}_L(x^\lambda,y) = 0
\quad , \qquad
\psi^{(0)}_R(x^\lambda,y) = e^{-A(y)} \psi^{(0)}_R(x^\lambda) ,
\neqno\chiralmode$$
as well as a series of paired massive left and right modes
$\psi^{(n)}_{L,R}(x^\lambda,y)$ for $n>0$.  Unlike the flat Kaluza-Klein
expansion, the zero mass mode does depend on the fifth coordinate through the
$e^{-A(y)}$ factor.

\newsec{Fermions and gauge fields in an oscillating background.}

The presence of ordinary fermion or gauge fields affects the evolution of
universe with rapid time oscillations.  The interaction of such fields with
the oscillating metric offers a route for the relaxation these oscillations. 
However, we shall see that the zero mass mode of the Kaluza-Klein expansion of
a massless five dimensional fermion does not couple directly to the
oscillating components of the metric, to leading order.  A decay into gauge
fields (or scalar fields) becomes the dominant channel for the relaxation of
the metric.

Returning to a metric with a dependence both on time and the extra dimension
$\oldmetric$, when $B(t)\not = C(t)$ this metric is not conformally flat. 
This feature becomes more apparent if we define new coordinates
$$\eta(t) \equiv \int^t e^{-{1\over 2}B(t')}\, dt'
\quad\hbox{and}\quad
r(y) \equiv \int^y e^{-{1\over 2}A(y')}\, dy' , \neqno\newreta$$
in terms of which the metric $\oldmetric$ becomes
$$ds^2 = e^{A(r)} e^{B(\eta)} \left[ - d\eta^2 + \delta_{ij}\, dx^i dx^j 
+ e^{C(\eta)-B(\eta)}\, dr^2 \right], \neqno\conmetric$$
where $A(r)$, $B(\eta)$ and $C(\eta)$ are understood to be $A(y(r))$,
$B(t(\eta))$ and $C(t(\eta))$.   

We now introduce a fermion field $\psi$ and an Abelian gauge field $A_a$
through the action\foot{Because of our choice for the metric's signature, we
have not included the usual factor of $i$ in the fermion action.  The $\gamma$
matrices are defined by $\Gamma^A = \{ i\gamma^\mu, \gamma_5 \}$, where
$\gamma^\mu$ and $\gamma_5$ are the standard $\gamma$ matrices, so that $\{
\Gamma^A,\Gamma^B \} = 2\eta^{AB}$ for $\eta^{AB}={\mathop{\rm
diag}}[-1,1,1,1,1]$.  Refer to 
\ref\threebrane{R.~Sundrum,
Phys.\ Rev.\ D {\bf 59}, 085009 (1999) [hep-ph/9805471].}  
or 
\ref\veltman{M.~Veltman, in {\it Methods of Field Theory\/}, Proceedings of 
the Les Houches Summer School, Les Houches, France, 1975, edited by R. Balian 
and J. Zinn-Justin, Les Houches Summer School Proceedings Vol. XXVIII 
(North-Holland, Amsterdam, 1976), p.~265.}  
for further details on fermions in a curved background.}
$$S_{\psi,A} = \int d^4xdr\, \sqrt{-g}\, \left\{ 
e^{\ a}_A \bar\psi\Gamma^A (D_a - ig A_a)\psi 
- {\textstyle{1\over 4}} g^{ab}g^{cd}F_{ac}F_{bd} \right\} , 
\neqno\psiAaction$$
where $F_{ab}$ is the gauge field strength.  We work in the limit in which
these fields do not significantly affect the background geometry.  In this
action, we have not assumed an orbifold structure in the extra dimension, but
as in the previous section one can be readily introduced.  The $e^{\ a}_A$ is
a vierbein which connects the curved coordinates with a locally Lorentzian
frame:
$$e^{\ a}_A e^{\ b}_B\, g_{ab} = \eta_{AB} . \neqno\vierbeins$$
The covariant derivative of a fermion,  
$$D_a = \partial_a + {\textstyle{1\over 2}} \omega_{aBC} \sigma^{BC} , 
\neqno\covderiv$$
involves the spin-connection defined by
$$\eqalign{
\omega_{aAB} &= 
{\textstyle{1\over 2}} e_A^{\ b} \left[ \partial_a e_{Bb}
 - \partial_b e_{Ba} \right] 
- {\textstyle{1\over 2}} e_B^{\ b} \left[ \partial_a e_{Ab}
 - \partial_b e_{Aa} \right] \cr
&\quad
- {\textstyle{1\over 2}} e_A^{\ c} e_B^{\ d} \left[ \partial_c e_{Dd} 
- \partial_d e_{Dc} \right] e^D_{\ a} \cr} \neqno\spincon$$
and $\sigma^{BC} = {1\over 4}[\Gamma^B,\Gamma^C]$.

The contributions from the spin connection are canceled by introducing a
rescaled fermion field defined by
$$\psi(x^\lambda, r) = e^{-A(r)}e^{-{3\over 4}B(\eta)}e^{-{1\over 4}C(\eta)} 
\Psi(x^\lambda, r) . \neqno\redefinepsi$$
The fermion-gauge field action then becomes
$$\eqalign{S_{\psi,A} &= 
\int d^4xdr\, \left\{ 
i\bar\Psi\gamma^\mu (\partial_\mu - ig A_\mu)\Psi 
+ e^{{1\over 2}(B-C)} \bar\Psi\gamma^5 (\partial_r - ig A_r)\Psi \right\} \cr 
&\quad 
+ \int d^4xdr\, e^{{1\over 2}(A+C)} \left\{ - {\textstyle{1\over 4}} 
\eta^{\mu\nu}\eta^{\lambda\rho} F_{\mu\lambda}F_{\nu\rho} 
- {\textstyle{1\over 4}} e^{(B-C)} \eta^{\mu\nu} F_{\mu r}F_{\nu r} 
\right\} ; \cr}
\neqno\PsiAaction$$
note that the first term no longer contains $A(r)$, $B(\eta)$ or $C(\eta)$.

\subsec{The Kaluza-Klein expansion.}

To investigate the effect of the oscillating metric in the long distance
limit, we expand the fermion and gauge fields in a tower of Kaluza-Klein
modes.  To simplify, we work in the axial gauge $A_r=0$.  The fermions are
expanded separately in left- and right-handed fields,
$$\Psi_{L,R}(x^\lambda,r) \equiv \half (1\mp\gamma_5)\Psi 
= \sum_{n=0}^\infty \Psi^{(n)}_{L,R}(x^\lambda) f^{(n)}_{L,R}(r) 
\neqno\KKfermions$$
where $f^{(n)}_{L,R}(r)$ satisfy
$$\partial_r f^{(n)}_L = m_n f^{(n)}_R 
\qquad
\partial_r f^{(n)}_R = - m_n f^{(n)}_L  \neqno\KKfmass$$
and obey the following orthogonality condition,
$$\int_{-r_c/2}^{r_c/2} dr\, f^{(m)*}_L(r) f^{(n)}_L(r) = 
\int_{-r_c/2}^{r_c/2} dr\, f^{(m)*}_R(r) f^{(n)}_R(r) = \delta^{mn} ,
 \neqno\KKfortho$$
where $r_c$ is the volume of the extra dimension.  Analogously, we expand the
gauge field
$$A_\mu(x^\lambda, r) = \sum_{n=0}^\infty A^{(n)}_\mu(x^\lambda) h^{(n)}(r) , 
\neqno\KKgauge$$
defining the masses of the modes through
$$\partial_r\left[ e^{{1\over 2}A}\partial_r h^{(n)} \right] 
= - M_n^2 e^{{1\over 2}A} h^{(n)} \neqno\KKAmass$$
and normalizing the states through 
$$\int_{-r_c/2}^{r_c/2} dr\, e^{{1\over 2}A(r)} h^{(m)}(r) h^{(n)}(r) 
= \delta^{mn} . \neqno\KKAortho$$
The fermion-gauge interaction will induce couplings among the various
Kaluza-Klein modes, 
$$G^{mnp}_{L,R} \equiv g \int_{-r_c/2}^{r_c/2} dr\, f^{(m)*}_{L,R}(r) 
f^{(n)}_{L,R}(r) h^{(p)}(r) . \neqno\KKcouplings$$
The effective action that appears in four dimensions as a result of
$\PsiAaction$ is thus
$$\eqalign{S_{\psi,A} &= 
\int d^4x\, \Big\{ \sum_{m,n} 
i\bar\Psi^{(m)}_L\gamma^\mu \Bigl( \delta^{mn}\partial_\mu 
- i\sum_p G_L^{mnp} A^{(p)}_\mu \Bigr) \Psi^{(n)}_L \cr
&\quad\qquad
+ \sum_{m,n} i\bar\Psi^{(m)}_R\gamma^\mu \Bigl( \delta^{mn}\partial_\mu -
 i\sum_p G_R^{mnp} A^{(p)}_\mu \Bigr) \Psi^{(n)}_R \cr
&\quad\qquad
+ e^{{1\over 2}(B-C)} \sum_n m_n \left( \bar\Psi^{(n)}_L\Psi^{(n)}_R 
+ \bar\Psi^{(n)}_R\Psi^{(n)}_L \right) \Bigr\} \cr 
&\quad 
+ \int d^4x\, e^{{1\over 2}C} \sum_n \left\{ - {\textstyle{1\over 4}} 
\eta^{\mu\nu}\eta^{\lambda\rho} F^{(n)}_{\mu\lambda}F^{(n)}_{\nu\rho} 
- {\textstyle{1\over 2}} e^{(B-C)} M_n^2\eta^{\mu\nu} A^{(n)}_\mu A^{(n)}_\nu 
\right\} . \cr}
\neqno\PsiAaction$$

At low energies, well below the Planck scale, only the dynamics of the
massless modes will be important in the effective theory.  The $r$-dependent
parts of the lowest-lying fermions are
$$f^{(0)}_L = f^{(0)}_R = r_c^{-1/2} $$
along with a factor of $e^{-A(r)}$ from $\redefinepsi$.  The couplings among
the massless modes simplify to
$$g_0 \equiv G^{000}_L = G^{000}_R = {g\over r_c} \int_{-r_c/2}^{r_c/2} dr\,
h^{(0)}(r) . \neqno\zerocoupling$$
The low energy effective action is thus
$$S_{\rm eff} = \int d^4x\, \, \Bigl\{ 
i\bar\Psi^{(0)} \gamma^\mu \bigl( \partial_\mu 
- i g_0 A^{(0)}_\mu \bigr) \Psi^{(0)} 
- {\textstyle{1\over 4}} e^{{1\over 2}C} F^{(0)}_{\mu\nu} F^{(0)\, \mu\nu} 
+ \cdots \Bigr\} . \neqno\zerofullaction$$

In this effective action, the gravitational oscillations do not couple
directly to the fermions, but rather to the gauge fields.  In the small
amplitude limit that we have assumed, the leading interaction is 
$${\cal L}_{\rm interaction} = - {\textstyle{1\over 8}} C(t) F^{(0)}_{\mu\nu}
F^{(0)\, \mu\nu} . \neqno\interaction$$
This coupling offers a natural channel for the decay of the oscillating
component of the metric.  Similarly, any fundamental scalar fields in the
theory would also allow the decay of the oscillating gravitational field.  The
kinetic energy term for the zero mass Kaluza-Klein mode of a massless five
dimensional scalar field has a prefactor of $e^{B+{1\over 2}C}$.  

If the fermion in $\zerofullaction$ has a small realistic mass $m$, either
through a fermion mass in the five dimensional action or if a massless fermion
subsequently develops a dynamical mass, a mass term would produce a coupling
between the fermion and the gravitational oscillations.  However, this term
would be suppressed by $m/M_5$ relative to $\interaction$.

\newsec{Ultra high energy cosmic rays.}

If some slowly decaying residual oscillations were to exist today, the gauge
field products of this decay would provide a possible source for the ultra
high energy cosmic rays.  Here we shall only focus on demonstrating that a
realistic cosmic ray spectrum can arise and establishing a rough limit on the
rate of the decay and shall not attempt to develop a detailed model that
produces this rate.  The dominant signal would likely result from a decay of
the oscillating background into a pair of gluons.  An initial gluon of energy
$E_0\sim M_{\rm Pl}\sim M_5$ will fragment into a high multiplicity jet of
particles with a wide range of energies.  For an initial photon pair the
secondaries are only those produced as the photons scatter from the
intergalactic radio background, and thus should not provide as strong a
constraint as gluon-gluon production.  Additionally, the presence of any gauge
interactions beyond the standard model could provide other decay channels.

The observed flux of ultra-high energy cosmic rays, assumed here to be
protons, can then set a bound on rate of production, or equivalently on the
rate of decay of the effective vacuum energy density, $\rho_{\rm vac}$.  We
estimate the flux of protons above some energy $E_{\scriptscriptstyle >}$ by
$$J_{\scriptscriptstyle >} \approx {\textstyle{1\over 4\pi}} 
\dot\rho_{\rm vac} E_0^{-1} \ell N_{\scriptscriptstyle >} .
\neqno\flux$$
$N_{\scriptscriptstyle >}$ is the number of protons with energy $E >
E_{\scriptscriptstyle >}$ in a jet produced by the initial gluon of energy
$E_0\sim M_{\rm Pl}$.  $\ell$ is the attenuation length of these protons due
to scattering from the cosmic microwave background radiation 
\ref\gzk{K.~Greisen,
Phys.\ Rev.\ Lett.\ {\bf 16}, 748 (1966) and G.~T.~Zatsepin and
V.~A.~Kuzmin,
JETP Lett.\  {\bf 4}, 78 (1966) 
[Pisma Zh.\ Eksp.\ Teor.\ Fiz.\  {\bf 4}, 114 (1966)].}.  
When $E_{\scriptscriptstyle >} \approx 10^{11}\, {\rm GeV}$ then $\ell$ is
approximately a few tens of Mpc 
\ref\ginzburg{V.~L.~Ginzburg, V.~A.~Dogiel, V.~S.~Berezinsky,
S.~V.~Bulanov and V.~S.~Ptuskin, {\it Astrophysics Of Cosmic Rays\/},
Amsterdam, Netherlands:  North-Holland (1990).}.   
This energy is just above the expected but not seen GZK cutoff $\gzk$, which
is exceeded here by high energy protons originating within a distance $\ell$
of us.

We shall obtain an estimate of $N_{\scriptscriptstyle >}$ from the
perturbative analysis of multiparticle production in jets based on the
modified leading log approximation ({\it e.g.\/}\ 
\ref\ochs{V.~A.~Khoze and W.~Ochs,
Int.\ J.\ Mod.\ Phys.\ A {\bf 12}, 2949 (1997) [hep-ph/9701421].}).\foot{In 
this approximation the main difference at leading order between a quark or
gluon jet is that the multiplicity of particles in the gluon jet is enhanced
by a factor of ${9\over 4}$ relative to the quark jet for the same initial jet
energy.  The mean particle energy of products will correspondingly be slightly
lower in the gluon jet.  In the following we have treated a gluon jet the same
as a quark jet.}  The resulting ``limiting spectrum'' is known but since the
initial gluon energy is so high, $\tau \equiv \ln(E_0/\Lambda_{\rm QCD}) \gg
1$, we may simplify further with a Gaussian representation.  We extract the
following results from $\ochs$ where $\xi \equiv \ln(E_0/E)$.
$${dN\over d\xi} \propto \exp \left( -{1\over 2} 
{(\xi-\overline{\xi})^2\over\sigma^2} \right)
\neqno\spec$$
$$\sigma   =  {\tau\over\sqrt{3z}} \left( 1 - {3\over 4z} \right)
\qquad
\overline{\xi} = \tau \left( {1\over 2} + \sqrt{{C\over\tau}} \right)
\neqno\cs$$
Here $z = \sqrt{16N_c\tau/b}$, $C = a^2/(16N_c b)$, $3b = 11N_c-2n_f$, and $3a
= 11N_c+2n_f/N_c^2$.  We also note that a Gaussian spectrum resembles the
results from Monte Carlo simulations 
\ref\bk{V.~Berezinsky and M.~Kachelriess,
Phys.\ Rev.\ D {\bf 63}, 034007 (2001) [hep-ph/0009053].}.   
For the total multiplicity ${\cal N} = \int_0^\infty (dN/d\xi)\, d\xi$ we use
the full ``limiting spectrum'' result,
$${\cal N}  =  \Gamma(B){\left( {z\over 2} \right)}^{1-B} I_{B+1}(z) ,
\neqno\limspec$$ 
where $B=a/b$.  This gives ${\cal N}\approx 7\times 10^5$.  Assuming that 5\%
of the energy from each initial gluon emerges as protons, these results imply
that $N_{\scriptscriptstyle >}\approx 3000$ for $E_{\scriptscriptstyle
>}=10^{11}\, {\rm GeV}$ and $n_f=6$.  This value for $N_{\scriptscriptstyle
>}$ triples when the ${\cal O}(\tau^{-1/2})$ corrections in $\cs$ are absent. 
Another indication of the sensitivity of the result to the approximation is
that no single value of $n_f$ is actually correct over the range of energies
involved, and $N_{\scriptscriptstyle >}$ ranges from 1400 to 6000 as $n_f$
ranges from 3 to 8.  This last result also shows how $N_{\scriptscriptstyle
>}$ can be strongly affected by new physics beyond the standard model.

The observed integrated flux of cosmic rays with $E>10^{11}\, {\rm GeV}$ is
approximately $2\times10^{-20}\, {\rm cm}^{-2}\, {\rm s}^{-1}\, {\rm sr}^{-1}$
\ref\nagano{M.~Nagano and A.~A.~Watson,
Rev.\ Mod.\ Phys.\  {\bf 72}, 689 (2000).},  
and so $\flux$ and $N_{\scriptscriptstyle >}\approx 3000$ implies that
$$\dot\rho_{\rm vac} < 3\times10^{-53}\, {\rm g}\, {\rm cm}^{-3}\, 
{\rm s}^{-1} . \neqno\rhodotlimit$$
Thus the current vacuum energy density of the universe, assumed to be about
two-thirds of the critical density, has a decay rate
$${\dot\rho_{\rm vac}\over \rho_{\rm vac}} 
< 10^{-6}\, {1\over t_{\rm universe}} ,\neqno\rhorhodot$$
where $t_{\rm universe}\approx 14$~Gyr is the age of the universe.

If this limit is saturated then we have a model for the ultra-high energy
cosmic rays.  The spectrum in $\spec$ implies a hard energy spectrum
$dN/dE\sim 10^{-\alpha}$ with $\alpha\approx 1.2$ at the relevant energies. 
The observed flux spectrum $dJ/dE$ is modified by the rapid decrease of the
attenuation length by almost a factor of 100 between
$4\times10^{10}<E<10^{11}\, {\rm GeV}$ (the GZK cutoff).  The frequently
plotted $E^3dJ/dE$ thus rises for $E<4\times10^{10}\, {\rm GeV}$, drops for
$4\times10^{10}<E<10^{11}\, {\rm GeV}$, and continues to rise for $E>10^{11}\,
{\rm GeV}$.  This behavior is a simple consequence of a hard initial spectrum
of protons produced uniformly throughout space, and is quite consistent with
the data above $10^{10}\, {\rm GeV}$.  Below $10^{10}\, {\rm GeV}$ the
observed flux rises much faster with decreasing energy, and some other source
for cosmic rays must dominate.

This picture is similar to models involving decay of supermassive long-lived
particles 
\ref\bkv{V.~Berezinsky, M.~Kachelriess and A.~Vilenkin,
Phys.\ Rev.\ Lett.\  {\bf 79}, 4302 (1997) [astro-ph/9708217];
V.~A.~Kuzmin and V.~A.~Rubakov,
Phys.\ Atom.\ Nucl.\  {\bf 61}, 1028 (1998)
[Yad.\ Fiz.\  {\bf 61}, 1122 (1998)] [astro-ph/9709187].}
\ref\smp{A.~G.~Doroshkevich and P.~D.~Naselsky,
astro-ph/0011320.}
\ref\FK{Z.~Fodor and S.~D.~Katz,
Phys.\ Rev.\ Lett.\  {\bf 86}, 3224 (2001) [hep-ph/0008204].}.   
The primary difference is that the supermassive particles tend to congregate
in the galactic halo, and this produces a galactic component to the signal in
addition to a possible extragalactic one.  But the galactic component would
not produce a dip in the spectrum, and it would produce some amount of large
scale anisotropy $\smp$ $\FK$.  In our picture the absence of a galactic
component is natural, although there remains the question of just how
spatially uniform the decay mechanism would be.

The production of cosmic gluon jets also ties in with another mechanism for
producing air showers above the GZK cutoff, this time involving jets produced
outside the $\ell^3$ volume.  This is because gluon jets contain neutrinos,
and these energetic cosmic neutrinos have some probability for producing
$Z$-bursts within the $\ell^3$ volume, as they travel toward us %
\ref\wlr{
T.~J.~Weiler,
Astropart.\ Phys.\  {\bf 11}, 303 (1999) [hep-ph/9710431];
Astrophys.\ J.\  {\bf 285}, 495 (1984);
Phys.\ Rev.\ Lett.\  {\bf 49}, 234 (1982);
E.~Roulet,
Phys.\ Rev.\ D {\bf 47}, 5247 (1993).}.   
In particular neutrinos with energy within $\delta
E/E_R=\Gamma_Z/M_Z\approx3\%$ of the $Z$-resonance energy $E_R=4\,({\rm
eV}/m_\nu )\times 10^{12}\,{\rm GeV}$ may annihilate with an enhanced cross
section on the nonrelativistic relic antineutrinos (and vice versa) to produce
the $Z$.  We estimate (with the same uncertainties as before) that the number
of neutrinos in this energy band, produced per gluon jet, ranges from 300 to
1100 as $E_R$ ranges from $10^{13}$ to $10^{11}\,{\rm GeV}$.  The probability
for such a neutrino to produce a $Z$-burst within the $\ell^3$ volume is in
the range $0.025\%$ to $1\%$, depending on the relic neutrino clustering
$\wlr$.  On the other hand there is an enhancement factor of about 100 for the
neutrino flux relative to the direct proton flux since the former originates
from the whole Hubble volume.  Each $Z$-burst produces a couple of protons
with typical energies $E_p\approx E_R/30$, and thus this mechanism produces
protons peaked in a fairly narrow energy range.  It may even be of interest if
this peak was somewhat below the GZK bound, where the protons-from-jets
mechanism was deficient.

\newsec{Conclusions.}

A warped, compact background space-time with a compact scalar field,
introduced to address the cosmological constant problem, does not present any
obstructions to including chiral fermions.  Since the compactness of the extra
dimension resulted from a periodic solution to the field equations, rather
than adding brane boundaries, the most appropriate approach is to give the
universe an $S^1/\hbox{\bbb Z}_2$ orbifold structure in this dimension.

A Kaluza-Klein expansion of massless fermion and gauge fields in a warped and
oscillating background reveals that the zero mass modes of the fermion do not
couple at leading order with the oscillating terms in the metric.  Thus the
gauge fields would most readily facilitate the decay of the oscillating
gravitational field.  

The decay of the oscillations in the metric should proceed rapidly in the
early universe, at least until the amplitude is sufficiently small that the
oscillatory effects are comparable to any de Sitter expansion or to effects
from any matter and radiation present, in which case the simple picture
developed in section 2 breaks down and the subsequent evolution is more
complex.  Yet it would be useful to determine when such oscillations could
coexist with a familiar cosmology in the $3+1$ dimensional effective theory
since we have seen that a small, slowly relaxing oscillation today could
provide a source for the observe ultra high energy cosmic rays.

\bigskip
\bigskip

\centerline{ {\bf Acknowledgments} } 

\noindent This work was supported in part by Natural Sciences and Engineering
Research Council of Canada.  The work of H.~C.\  was also supported in part
by the Department of Energy under grant DOE-ER-40682-143.

\listrefs 

\bye